# Smart Navigation System for Parking Assignment at Large Events: Incorporating Heterogeneous Driver Characteristics

Xi Cheng[1*], Tong Liu[2], Gaofeng Su[3], Chang Che[4], Chen Zhu[5], Ke Liu[3], Binze Cai[6] and Xin Hu[7]

[1] Cornell University, NY, USA
[2] University of Illinois at Urbana-Champaign, IL, USA
[3] University of California at Berkeley, CA, USA
[4] The George Washington University, DC, USA
[5] Tsinghua University, Beijing, China
[6] Georgia Institute of Technology, GA, USA
[7] University of Michigan, Ann Arbor, MI, USA

[1] xc557@cornell.edu
[2] tongl5@illinois.edu
[3] koufongso, liuke126@berkeley.edu
[4] cche57@gwmail.gwu.edu
[5] zhuchen9505@foxmail.com
[6] binzecai@gmail.com
[7] hsinhu@umich.edu
*corresponding author

**Abstract.** Parking challenges escalate significantly during large events such as concerts and sports games, yet few studies address dynamic parking lot assignments in these occasions. This paper introduces a smart navigation system designed to optimize parking assignments efficiently during major events, employing a mixed search algorithm that considers diverse drivers characteristics. We validated our system through simulations conducted in Berkeley, CA during the "Big Game" showcasing the advantages of our novel parking assignment approach.

**Keywords:** Smart Parking Systems, Parking Assignment, Diverse Driver Behavior.

## 1. Introduction

With increasing urbanization, cities encounter growing challenges of traffic congestion and parking shortages. Limited space and high construction costs for new parking facilities exacerbate the problem, particularly in dense metropolitan areas. These issues intensify during large events or in central business districts. Researchers found that about 30% of cars on the road in the downtown area of major cities [1], seemed to be cruising for parking spots, which took an average of 7.8 min [2].

Researchers have introduced various parking assignment methods, including parking guidance systems, to simulate driver behavior and assess the impact of these systems. For instance, parking guidance systems proposed by Shin and Jun (2014) [3], aiming to reduce driver cruising time by directing them to available spaces. Simulation experiments were conducted to evaluate the effectiveness of these systems, with a base case assuming drivers head to the nearest parking facility without guidance. Furthermore, with the rise of smart devices and intelligent infrastructure systems, intelligent parking guidance systems can now provide real-time information on available parking spaces and direct drivers accordingly. For instance, Rehena et al. (2018) [4] proposed a multi-criteria parking reservation algorithm considering user preferences, such as distance, cost, and availability.

However, the impact of heterogeneous driver characteristics on parking guidance systems has been largely overlooked in the existing literature. To bridge this gap, this paper proposes a comprehensive framework for parking optimization and evaluates the effectiveness of a parking guidance algorithm that accounts for diverse driver behaviors. Optimization models have been widely applied in traffic systems, including freight transportation [5], [6], public transit [7], [8], carsharing services [9], and electric vehicles [10]. The framework integrates factors such as parking lot proximity and walking distance to more accurately simulate real-world conditions by assessing various search routes and strategies. By leveraging these methods, each vehicle is dynamically assigned an optimal search route, ensuring a more personalized and efficient parking experience.

The paper is structured as follows: Section 2 details the proposed parking guidance algorithm for assigning vehicles to the most suitable parking lots. In Section 3, we evaluate the benefits of the algorithm using baseline comparisons, including strategies such as searching for parking near the destination, starting from the current location, and random nearby lots. Section 4 presents and discusses the results, demonstrating the feasibility and effectiveness of the algorithm. Section 5 provides the conclusion and further discussion.

## 2. Methodology

### 2.1. Parking assignment optimization method

In urban areas with high parking demand, efficient vehicle assignment to parking facilities is crucial to reducing vehicle miles traveled (VMT), alleviating traffic congestion from parking searches, and mitigating environmental impacts. This paper presents an optimization-based parking assignment method designed to minimize drivers' search time for parking spaces. The method targets minimizing time spent from entry points to parking lots, the search for parking spots, and walking time from the lot to the destination. The optimization problem incorporates several constraints: each vehicle must be assigned to a specific parking lot, parking lot capacities must not be exceeded, and binary constraints govern assignment decisions. The problem is formulated as follows:

$$\min_{x_{ik}} \sum_{i=1}^{m} \sum_{k=1}^{K} x_{ik} \cdot (TD_{ik} + TS_{ik} + TW_i)$$

$$s.t. \quad \sum_{i=1}^{m} \sum_{k=1}^{K} x_{ik} = K,$$

$$\sum_{k=1}^{K} x_{ik} = VolLot_i,$$

$$x_{ik} = \{0,1\},$$

where $TD_{ik}$ is the driving time from the entry points, which is appointed to the vehicle $k$ randomly in advance, to the parking spot $i$ of the vehicle $k$. $TS_{ik}$ is the search time in parking lot $i$ of vehicle $k$, which is the duration from entering the parking lot to exiting. $TW_i$ is the walking time from parking lot $i$ to the destination. $x_{ik}$ is 1 if vehicle $k$ is assigned to lot $i$. Assuming all drivers park in the space at single-entry parking lot in sequence, as shown in Figure 1, $TS_{ik}$ can be estimated using the following equations:

$$TS_{ik} = \left(\frac{W \cdot VolLot_i \cdot O_i}{2v_c} + \frac{W \cdot VolLot_i \cdot O_i}{2v_w} + t_{stop}\right) + (N_i - 1) \cdot \left(\frac{R_{iL}}{v_{ud}} \cdot C_{iN} \cdot O_{iN} + t_{turn}\right),$$

where $O_i$ and $N_i$ represent the occupancy and number of floors of parking lot $i$. $O_{iN}$ and $C_{iN}$ represent the occupancy and the capacity of $N^{th}$ floor of parking lot $i$. $VolLot_i$ represent the capacity of parking lot $i$. $R_{iL}$ denotes length of a ramp of parking lot $i$. $W$ denotes the width of a parking spot. $v_c$ and $v_w$ denote the average cruising speed and walking speed. $v_{ud}$ is average speed on the ramp. $t_{stop}$ and $t_{turn}$ denote the average parking time and U-turn time between ramps.

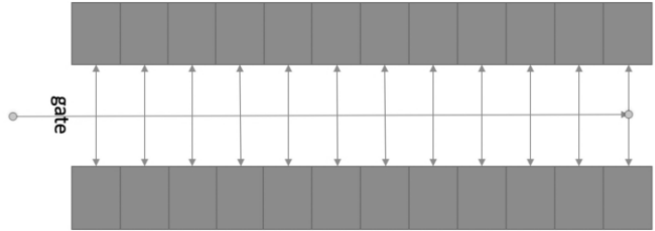

**Figure 1.** Layout of inner parking lots

*2.2. Parking choices with driver heterogeneity*
Previous studies have evaluated parking assignment methods by simulating real-life scenarios under the assumption that all drivers follow a uniform strategy. However, in real-world scenarios, parking choices result from interaction between drivers' preferences, the immediate availability of spaces, and current traffic conditions [4]. This paper introduces a method for parking search that simulates situations in which drivers are unfamiliar with parking options near their destination. Furthermore, this paper integrates multiple strategies by assigning different weights to each, aiming for a more accurate reflection of real-world conditions.

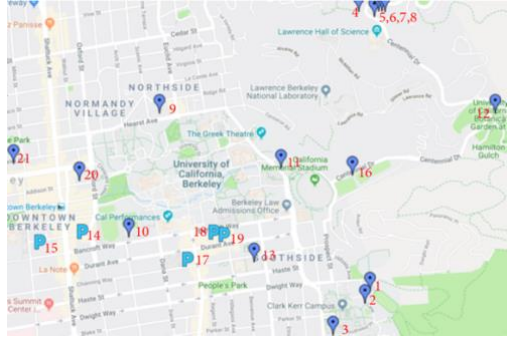

**Figure 2.** Region used in case study around California memorial stadium.

The study area is defined as a rectangular region encompassing all parking lots near the destination, as shown in Figure 2. Vehicles are categorized based on different search strategies and enter the region through m traffic links, following a specified arrival time distribution. $AT_i$, the actual arrival time of vehicle $i$ is expressed as:

$$AT_i = ET_i + NS_i$$

where $ET_i$ is expected arrival time of vehicle $i$, which conforms to Poisson distribution. $NS_i$ represent the noise of arrival time of vehicle $i$, which conforms to normal distribution. In our model, vehicles are assigned to distinct search groups, each using a specific strategy for selecting parking lots. If a vehicle encounters a full lot, it is redirected to the next most suitable option. However, drivers exhibit limited tolerance for repeatedly encountering full lots and may opt for more distant parking, which is less convenient. To quantify this tolerance, we model the probability of a driver abandoning their search as a function of time spent searching, using a Gamma distribution to capture variability in driver behavior. Prolonged search times can result in drivers leaving the designated parking area, increasing congestion, emissions, and time waste. This model enables the simulation and analysis of parking availability and search strategies' impact on urban congestion and environmental outcomes. By understanding the limits of driver patience, we can refine parking assignment algorithms to mitigate negative effects and enhance the efficiency of urban transportation systems.

## 3. Experimental setup

A significant influx of American football fans drives to the annual game between UC Berkeley and Stanford, held at California Memorial Stadium. Despite some fans' familiarity with Berkeley, the arrival of many new visitors complicates parking decisions for all. In such scenarios, even local drivers struggle to identify optimal parking spots due to changing conditions. To address this, our model treats all drivers as unfamiliar with the area to simplify guidance to suitable parking spaces, enhancing safety by reducing reliance on maps which can distract drivers.

Comprehensive data on the parking infrastructure around California Memorial Stadium has been collected. This includes the exact locations and capacities of all visitor parking lots. The dataset reveals that there are 21 parking lots within a designated rectangular region around the stadium, providing a total of 3,992 parking spaces. Furthermore, we assume the parking demand equals the supply, utilizing all 3,992 spaces. Our parking assignment algorithm optimally allocates vehicles to specific spaces, minimizing total travel time from entry points to parking spots. The system models uniform vehicle arrivals starting two hours before the event across 12 access streets. Furthermore, we simulate realistic scenarios when all vehicles attempt to enter simultaneously. We conducted comparative analyses with established models from prior studies. To effectively manage parking during the event, vehicles are divided into four search groups, each employing a distinct strategy for selecting parking lots., which have the same weight.

- Group 1: Prioritizes finding the nearest available parking lot upon entering the region, aiming to minimize driving distance from entry points.
- Group 2: Seeks the nearest parking lot while avoiding areas close to the stadium to reduce congestion and distribute parking usage.
- Group 3: Focuses on minimizing walking distance to the stadium, prioritizing convenience for attendees.
- Group 4: Optimizes parking by minimizing total travel time, with both driving and walking distances.

Furthermore, to facilitate the computation of these distances and the subsequent travel times for each vehicle, we converted geographic coordinates into Cartesian coordinates using Miller projections [11]:

$$x' = M_x(lon) = lon$$
$$y' = M_y(lat) = 1.25 \cdot \ln\left(tan\left(\frac{\pi}{4} + 0.4 \cdot lat\right)\right)$$
$$L = 6381372 \cdot \pi \cdot 2$$
$$x = \frac{L}{2} + \frac{L}{2 \cdot \pi} \cdot x'$$
$$y = \frac{L}{4} - \frac{L}{4 \cdot 2.3} \cdot y'$$

And in this paper, we assume the sideways taken into consider are perpendicular with each other (see Figure 3). Therefore, the Manhattan distance between parking lot $i$ and parking lot $j$ or the destination is going to be equal to $|x_i - x_j| + |y_i - y_j|$, where $x_i, y_i, x_j, y_j$ are the horizontal and vertical coordinates of the parking lot $i$ and parking lot $j$. The optimization problem is framed for a rush hour period from 10 a.m. to 12 p.m. To simplify the problem, we divide this total time into 12 segments using 10-minute intervals.

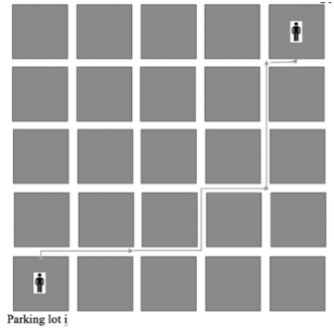 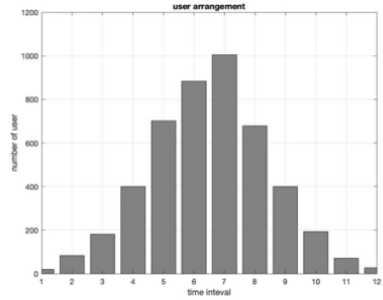

**Figure 3.** Distance calculation demonstration.

**Figure 4.** Distribution of arrival vehicles arrangement

While the arrival time for each individual can be arbitrarily assigned, it's more realistic to simulate a scenario where most people arrive around 11 a.m. Specifically, we use a Poisson distribution model to simulate the expected arrival time for each individual. The distribution of vehicle arrival times is illustrated in Figure 4.

## 4. Results

Utilizing the previously discussed arrival times and parking strategies, we performed simulations to model the parking process in the designated case region. Our primary analysis compared group-based parking strategies with the optimal assignment results. The simulations indicate that the average rerouting time per vehicle was 4.7 minutes, serving as a key metric for evaluating the efficiency of our parking management system. To validate the effectiveness of the proposed approach, we compared our results with baseline scenarios, as shown in Table 1. Our model shows a significant reduction in average rerouting time, highlighting the time-saving potential of our assignment method.

**Table 1.** Comparison of base cases of average rerouted time for each driver (min)

| Comparisons | Shin's method | Rehena's method | Our model |
|---|---|---|---|
| Time (min) | 14.6 | 6.1 | 4.7 |

To account for the inherent uncertainty and randomness in vehicle arrival times, we conducted multiple simulations to ensure the robustness of our findings. Figure 5 illustrates the relationship between the number of simulations and the average rerouting time, demonstrating the robustness of the model in handling variability and maintaining consistent performance across different scenarios.

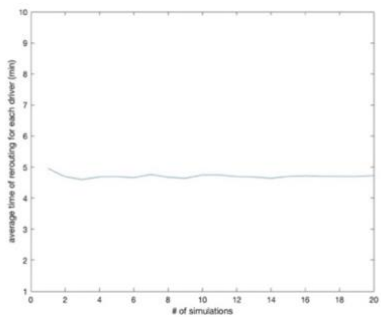 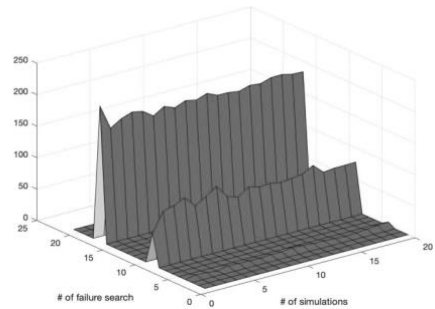

**Figure 5.** Relationship between average rerouting time and number of simulations.

**Figure 6.** Failure search with different parking lot and number of simulations.

A more robust criterion for comparing the effectiveness of different parking guidance methods is the number of failed searches. Figure 6 illustrates the relationship between the number of failed searches across various parking lots and the number of runs. After approximately 10 simulations, both the average rerouting time and the number of failed searches stabilize, indicating that 10 simulations are sufficient to establish a reliable baseline for evaluating parking performance.

**5. Conclusions**

This paper presents an enhanced parking assignment method that strategically allocates vehicles to parking structures, incorporating a simulation of real-life parking scenarios. Our optimization model reduces rerouting times by accounting for the heterogeneous characteristics of drivers, offering a significant improvement over traditional approaches. Unlike prior studies that rely on unvalidated theoretical models, our method uses real parking data to more accurately estimate search times, leading to a more realistic assessment of parking management strategies. Additionally, we developed a simulated baseline that mirrors real-life conditions, providing a strong basis for evaluating the effectiveness of our optimization. While our simulation marks a substantial improvement, it does not fully capture the complexity of actual driver behavior, highlighting the need for future research to refine these models and integrate emerging technologies to better reflect real-world dynamics.